\documentclass[manuscript]{aastex}

\begin{document}

\title{Optical and $\gamma$-Ray Variability Behaviors of 3C 454.3 from 2006 to 2011}

\author{Xu-Liang Fan\altaffilmark{1}, Shao-Kun Li\altaffilmark{2,3}, Neng-Hui Liao\altaffilmark{4}, Liang Chen\altaffilmark{5,6}, Hong-Tao Liu\altaffilmark{2,3}, Kai-Xing Lu\altaffilmark{2,3}, Da-Hai Yan\altaffilmark{2,3}, Rui-Yu Zhang\altaffilmark{5,6}, Qian Guo\altaffilmark{2,3,6}, Qingwen Wu\altaffilmark{1}, and Jin-Ming Bai\altaffilmark{2,3}}

\altaffiltext{1}{School of Physics, Huazhong University of Science and Technology, Wuhan 430074, China, fanxl@hust.edu.cn, qwwu@hust.edu.cn}
\altaffiltext{2}{Yunnan Observatories, Chinese Academy of Sciences, Kunming 650011, China, baijinming@ynao.ac.cn}
\altaffiltext{3}{Key Laboratory for the Structure and Evolution of Celestial Objects, Chinese Academy of Sciences, Kunming 650011, China}
\altaffiltext{4}{Key Laboratory of Dark Matter and Space Astronomy, Purple Mountain Observatory, Chinese Academy of Sciences, Nanjing 210008, China}
\altaffiltext{5}{Key Laboratory for Research in Galaxies and Cosmology, Shanghai Astronomical Observatory, Chinese Academy of Sciences, 80 Nandan Road, Shanghai 200030, China}
\altaffiltext{6}{University of Chinese Academy of Sciences, Beijing 100049, China}

\begin{abstract}
 We present our photometric monitoring of a flat spectrum radio quasar (FSRQ) 3C 454.3 at Yunnan observatories from 2006 to 2011. We find that the optical color of 3C 454.3 shows obvious redder-when-brighter trend, which reaches a saturation stage when the source is brighter than 15.15 mag at V band. We perform a simulation with multiple values of disk luminosity and spectral index to reproduce the magnitude-color diagram. The results show that the contamination caused by the disk radiation alone is difficult to produce the observed color variability. The variability properties during the outburst in December 2009 are also compared with $\gamma$-ray data derived from Fermi $\gamma$-ray space telescope. The flux variation of these two bands follow a linear relation with $F_{\gamma} \propto F_R^{1.14\pm0.07}$, which provides an observational evidence for external Compton process in 3C 454.3. Meanwhile, this flux correlation indicates that electron injection is the main mechanism for variability origin. We also explore the variation of the flux ratio $F_{\gamma}/F_R$ and the detailed structures in the lightcurves, and discuss some possible origins for the detailed variability behaviors.
\end{abstract}


\section{Introduction}
Blazars are the most extreme subclass of active galactic nuclei. They show dramatic variability throughout the entire electromagnetic spectrum. Although the mechanisms producing variability remain unclear, multi-wavelength variability is frequently used to constrain the radiation mechanisms and emission regions of blazars~\citep{2011ApJ...726L..13A, 2011MNRAS.414..155L, 2014ApJ...783...83L}. Recently, several attempts were performed to explore common characteristics in variability itself~\citep{2014ApJ...791...21F, 2016MNRAS.455L...1W, 2017PASP..129a4101L, 2017ApJ...843...81O}.

3C 454.3 (2251+168) at z=0.859 is a typical flat spectrum radio quasar (FSRQ). FSRQ is a subclass of blazar, which shows prominent emission lines and thermal radiation from accretion disk. Their $\gamma$-ray emission is usually explained as external Compton (EC) radiation of the relativistic electrons, which also produce the synchrotron radiation at low energy band. However, this view still lacks direct observational evidence~\citep{2011ApJ...735..108C, 2012ApJ...752L...4M, 2016RAA....16..173F}. 3C 454.3 started an extreme activity in 2005~\citep{2006A&A...453..817V}. Then it kept active and generated several violent outbursts in the last decade throughout multi-wavelength (\citealt{2008A&A...491..755R, 2009ApJ...699..817A, 2010ApJ...721.1383A, 2010ApJ...716L.170P, 2011A&A...534A..87R, 2011ApJ...733L..26A, 2012PASJ...64...58S, 2012ApJ...758...72W, 2013ApJ...763L..36L, 2013ApJ...773..147J, 2017MNRAS.464.2046K}, also see the $\gamma$-ray lightcurve of Fermi Gamma-ray Space Telescope at the website of Fermi/LAT monitored sources~\footnote{https://fermi.gsfc.nasa.gov/ssc/data/access/lat/msl\_lc/source/3C\_454d3}). The IR, optical, UV and $\gamma$-ray outbursts occur almost simultaneously while the variation of millimeter (mm) radio flux generally lags the high frequencies (\citealt{2008A&A...491..755R, 2011A&A...534A..87R, 2013ApJ...773..147J}, simultaneous mm wave flares were also observed by~\citealt{2012ApJ...758...72W}). The X-ray variability is more complicated. In the 2008 outburst, no corresponding X-ray variability was observed by \textit{Swift}~\citep{2009ApJ...697L..81B}, while a slightly decayed X-ray flare was observed during the 2009 outburst~\citep{2010ApJ...716L.170P, 2011A&A...534A..87R}. In 2010 outburst, the X-ray flare seemed simultaneous with optical~\citep{2012ApJ...758...72W, 2013ApJ...773..147J}. In addition, the low frequency radio emission (lower than 8 GHz) turned into a low activity stage when the high frequencies became active~\citep{2006A&A...453..817V, 2008A&A...491..755R}.

The outburst of 3C 454.3 in December 2009 had been extensively explored by many authors at various wavelengths~\citep{2010ApJ...716L.170P, 2011A&A...534A..87R, 2011MNRAS.410..368B, 2012PASJ...64...58S, 2013ApJ...773..147J}. However, there are still some observations that have not been clearly understood, especially about the radiation mechanism of $\gamma$-ray emission and the variability origins.~\citet{2011MNRAS.410..368B} found that the $\gamma$-ray flux varied quadratically with the optical flux, which is inconsistent with the prediction of commonly assumed EC process in FSRQs.~\citet{2011A&A...534A..87R} found that the major optical flare corresponded to the minor flare at $\gamma$-ray band, which is difficult to understand under the one-zone leptonic model.~\citet{2013ApJ...773..147J} performed detailed analyses on the multi-wavelength variability and the polarization variation of 3C 454.3, including the outburst in 2009. They presented several theoretical models beyond one-zone leptonic model to interpret the multi-wavelength behaviors, while it is still unclear which model is better.

Obvious redder-when-brighter trend at relatively faint state, following a saturation effect at bright state, is the typical feature for the optical color variability of 3C 454.3~\citep{2008A&A...491..755R, 2015NewA...36...19Z}. The redder-when-brighter trend, which is often observed in FSRQs~\citep{2006A&A...450...39G, 2012ApJ...756...13B}, is usually explained as the contribution from less variable, bluer accretion disk to the strongly variable, redder jet emission. In addition, more complicated color behaviors of 3C 454.3 have also been observed at short time scale, which may indicate multiple emission components in this source~\citep{2011A&A...531A..90Z}.

In this paper, we present our optical monitoring data of 3C 454.3 from 2006 to 2011. Combined with the data from Whole Earth Blazar Telescope (WEBT) campaign, we investigate the origin of the optical color variability of 3C 454.3, and re-analyze its optical and $\gamma$-ray variability during the Dec. 2009 outburst. This paper is organized as follows. In Section 2 we present our optical multi-color monitoring of 3C 454.3 and the data reduction for optical and Fermi/LAT $\gamma$-ray data. Section 3.1 presents the explorations for optical color variability. In section 3.2 we investigate the correlation between optical and $\gamma$-ray flux. The variation of the flux ratio between $\gamma$-ray and optical is discussed in section 3.3. In section 3.4 we compare the detailed lightcurve between optical and $\gamma$-ray. Our main conclusions are summarized in Section 4.

\section{Observations and Data Reduction}
\subsection{Optical monitoring}
Our photometric monitoring of 3C 454.3 was performed with Lijiang 2.4 m telescope~\footnote{http://www.gmg.org.cn/english/} and Kunming 1 m telescope at Yunnan observatories from September 2006 to November 2011. A PIVersArry 1300B CCD with $1340 \times 1300$ pixels was used on Lijiang 2.4 m telescope before 2011, which covered a field of view of $4 \arcmin48 \arcsec \times 4\arcmin40 \arcsec$. After 2011, the Yunnan Faint Object Spectrograph and Camera (YFOSC) began to apply for most photometric and spectroscopy observations on Lijiang 2.4 m telescope. YFOSC has a field of view of about $10' \times 10'$ and $2148 \times 2200$ pixels for photometric observation. Each pixel corresponds to a sky angle of $0.283 \arcsec$. Before July 30th, 2008, a PI 1024TKB CCD with 1024 $\times$ 1024 pixels was equipped at Kunming 1.02 m telescope. The field of view is $6.5' \times 6.5'$. After that, it was updated to an Andor DW436 CCD with $2048 \times 2048$ pixels. The field of view is $7.8' \times 7.8'$. The photometric observations were performed with the standard Johnson/Cousins BVRI filters. Different exposure times were applied to match various seeing and weather conditions.

\begin{figure}
\epsscale{1.}
\plotone{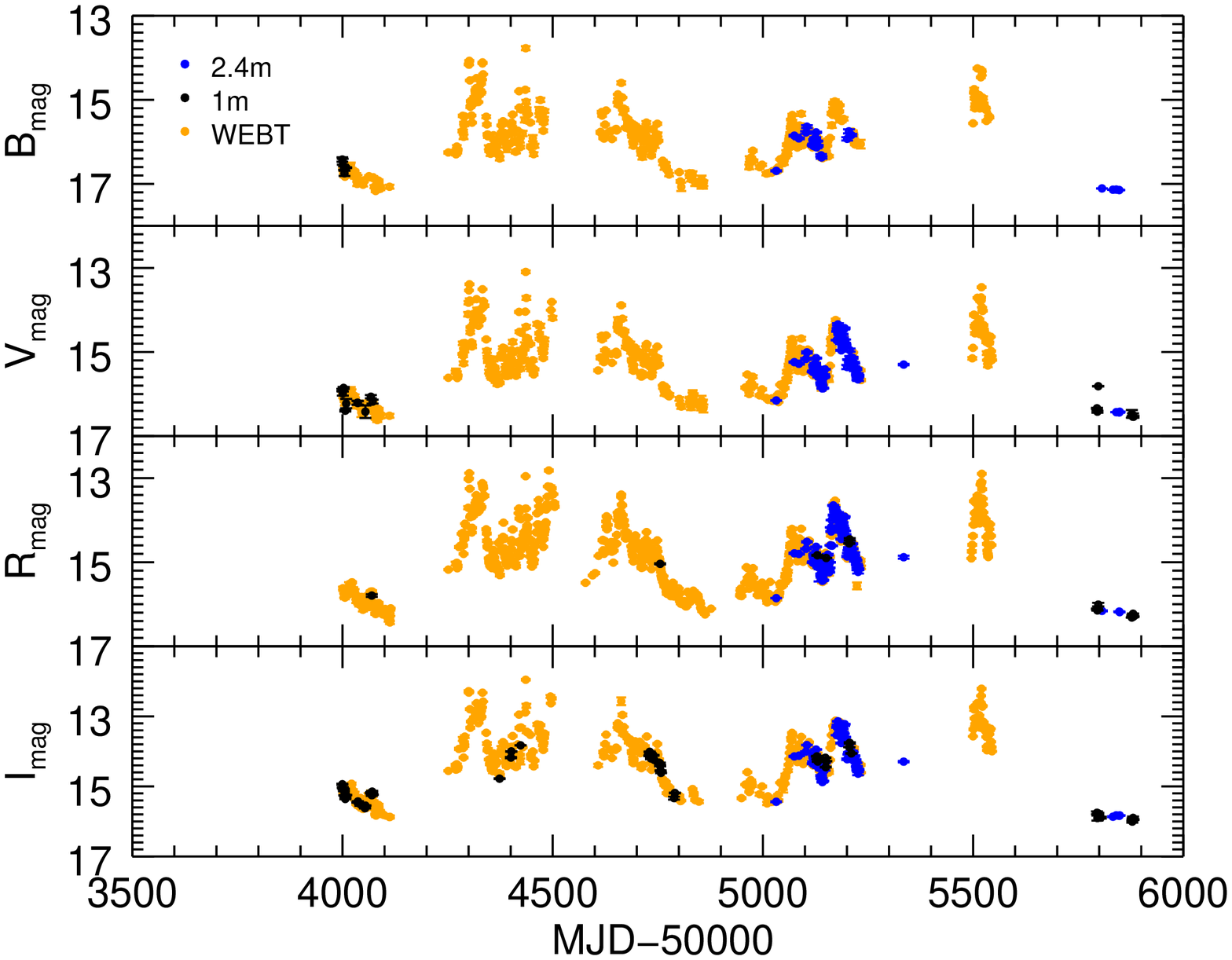}
\caption{The lightcurves of daily averaged magnitude for optical multi-band observations. From top to bottom are B, V, R and I band, respectively. The blue points manifest the data observed by Lijiang 2.4 m telescope. The black points represent the data observed by Kunming 1 m telescope. The orange points represent the data taken from WEBT. \label{lc}}
\end{figure}

The photometric data are reduced with the standard IRAF procedure, including the bias substraction and the flat field correction. Then the aperture photometry is performed with the \textit{APPHOT} package. The differential photometry is applied with the comparison star D in the finding chart~\footnote{ https://www.lsw.uni-heidelberg.de/projects/extragalactic/charts/2251+158.html}. The observational uncertainty of each night is estimated by the standard deviation of the magnitude difference between comparison stars D and E. For the B band data, the magnitude differences between comparison stars D and E have large deviations. ~\citet{2012ApJ...756...13B} also mentioned the same problem. They got the new B band magnitudes of the comparison stars based on the SMARTS data sets. Thus in our work, the differential photometry for B band is based on the finding chart of SMARTS~\footnote{http://www.astro.yale.edu/smarts/glast/fc3C454.3.php}. For multiple exposures during individual night, we average their magnitudes. The daily averaged magnitudes are presented in Table~\ref{data}. Multi-band lightcurves are presented in Figure~\ref{lc}. For the analyses in this paper, the galactic extinction is corrected according to the values from NED ($A_B = 0.462$, $A_V = 0.355$, $A_R = 0.286$, $A_I = 0.208$) which are based on the dust map of~\citet{1998ApJ...500..525S}. For the December 2009 outburst (from MJD 55120 to 55220), our observations get nearly one data point per night at R band, which provide a good opportunity to study the correlated variability between optical and $\gamma$-ray bands. In order to construct more continuous multi-band lightcurves between 2006 and 2011, we also include the data from WEBT at the same period (from MJD 54000 to 56000) in our analyses~\citep{2011A&A...534A..87R, 2011ApJ...736L..38V}.

\subsection{Fermi/LAT $\gamma$-ray data reduction}
The $\gamma$-ray data of Fermi/LAT for 3C 454.3 are reduced with the Fermi ScienceTools v10r0p5 (more details about the Fermi/LAT data reduction can be found in~\citealt{2016ApJS..226...17L}). The newest Pass 8 data of 3C 454.3 are downloaded from the Fermi data server. The data from 1 September 2009 (MET 273456002) to 1 February 2010 (MET 286675202), in the energy range 0.1 --- 300 GeV, and in the region of interest (ROI) $10^\circ$ are selected with \textit{gtselect}. After the standard preprocess threads, the unbinned likelihood analysis is applied to extract the flux and spectrum with \textit{gtlike}. All 3FGL sources within $20^\circ$ around the target are included in the likelihood analysis. Considering the rare photons on daily timescale, a simple powerlaw model is used to construct the 1-day bin lightcurve. An overall fit is performed firstly for the whole time range. Then the spectral parameters of the background sources are fixed to the values of the overall fitting. The 1-day bin lightcurves of the integral flux and spectral index are plotted in Figure~\ref{lc3} (e) and (f), respectively.

\section{Results and Discussions}
\subsection{Optical color variability}
\begin{figure}
\begin{center}
\includegraphics[angle=0,scale=.35]{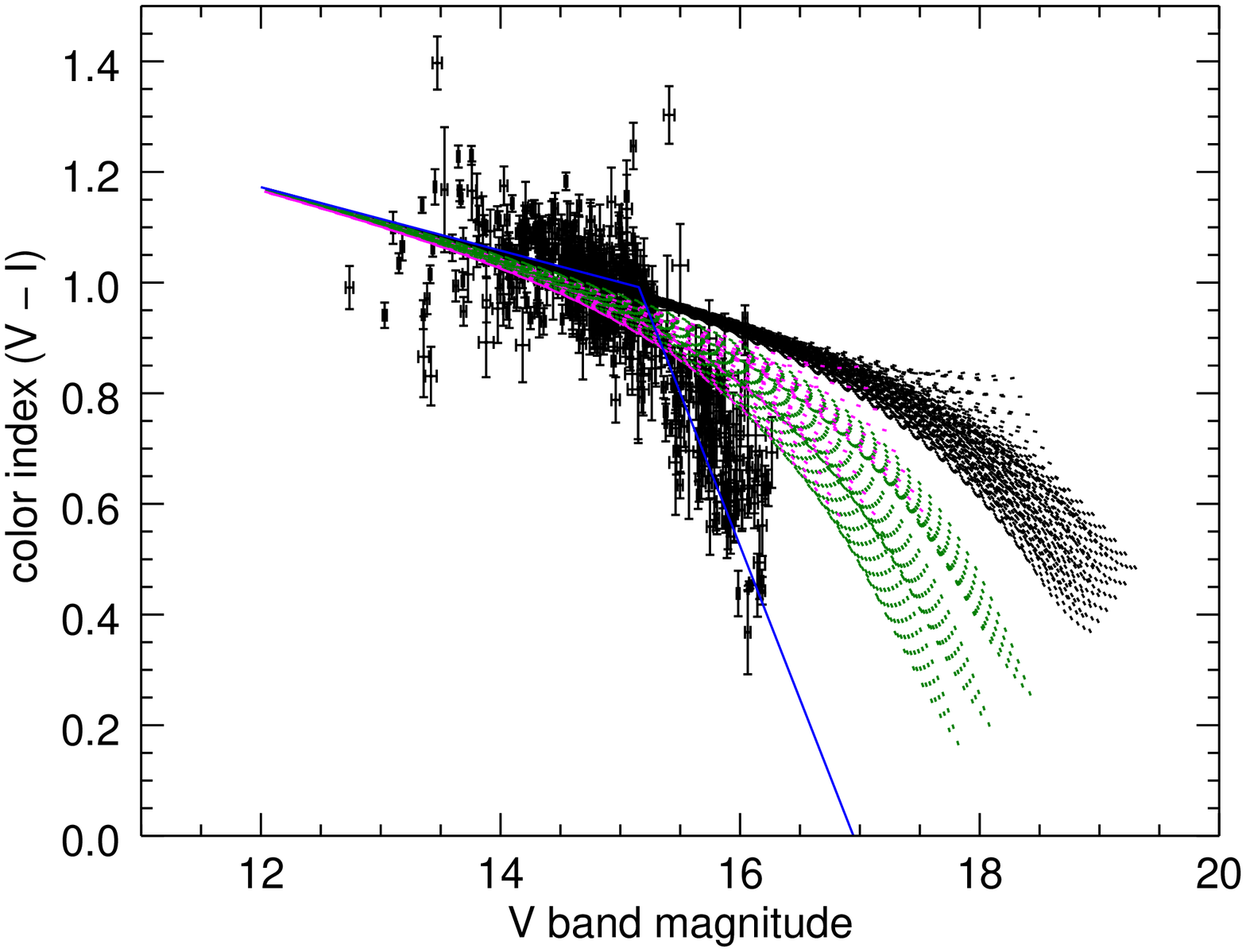}
\includegraphics[angle=0,scale=.35]{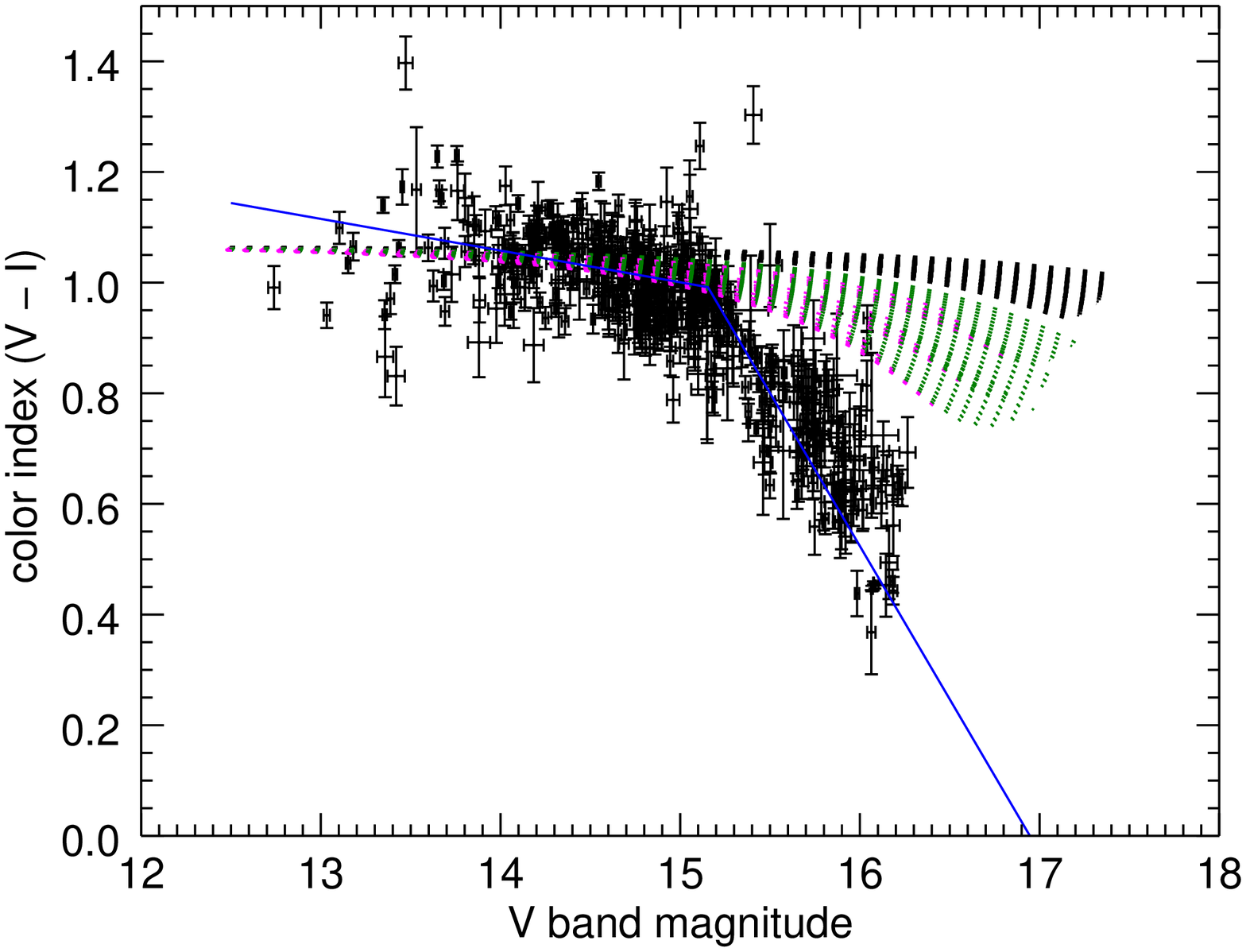}
\caption{The color index V - I versus V band magnitude. The blue solid line shows the best fit of two linear slopes. The black dot lines are the simulated results for disk luminosity fainter than $10^{46}$ erg s$^{-1}$. The green dot lines show the results for disk luminosity brighter than $10^{46}$ erg s$^{-1}$ and the spectral index less than 0.3. The magenta dot lines show the results for disk luminosity brighter than $10^{46}$ erg s$^{-1}$ and the spectral index larger than 0.3. Left panel is the results based on the relation between magnitude and color index from equation~\ref{satci}, right panel is the results based on a constant color index. See the text for details. \label{ci}}
\end{center}
\end{figure}

The color index V - I shows similar variation trend with the optical and $\gamma$-ray brightness (Figure~\ref{lc3}, panel g). which means that the source gets redder when the source brightens. The color indices V - I versus the V band magnitudes are plotted in Figure~\ref{ci}. An obvious redder-when-brighter trend is also shown, where the slope gets flatter when the object is brighter than a critical magnitude. We fit the data with a piecewise function with two linear slopes. The fitting result is overplotted in Figure~\ref{ci}, with the transition point at V$_{\rm mag}$ = 15.15. When the source is fainter than 15.15 mag, it shows a clear redder-when-brighter trend with
\begin{equation}
 \rm ci = -0.55~V_{mag}+9.38.
\end{equation}
When the source is brighter than 15.15 mag, it reaches a saturation stage with
\begin{equation}
 \rm ci = -0.06~V_{mag}+1.86.
\label{satci}
\end{equation}

The similar trends have also been found by~\citet{2008A&A...491..755R} and ~\citet{2015NewA...36...19Z}. This two-stage trend is usually explained as the existence of the thermal component from the accretion disk~\citep{2008A&A...491..755R, 2017ApJ...844..107I}. The saturation effect implies that the optical emission is dominated by jet radiation. Thus the transition magnitude gives an upper limit of the disk flux 3.08 mJy (with the zero point 3636 Jy,~\citealt{1998A&A...333..231B}), which corresponds to $1.14\times10^{47}$ erg s$^{-1}$ at 2931.7 \AA. This value is much larger than the results derived from the \textit{GALEX} observation at 1350 \AA~\citep[$4.0 \times 10^{46}$ erg s$^{-1}$, which corresponds to 16.73 mag at V band if the powerlaw slope $\alpha = 0.5$~\footnote{$F \propto \nu^{-\alpha}$} is assumed]{2011MNRAS.410..368B} and the value estimated from the BLR luminosity~\citep[$3.33 \times 10^{46} $ erg s$^{-1}$]{2012MNRAS.421.1764S}.

In order to verify whether such low contribution from disk can produce the observed color variability, we attempt to reproduce the magnitude-color relation for the first time. Firstly, the color indices of the jet emission between 0.5 and 1.5 (spectral index between 0.14 and 2.55) are inputted. The V band magnitudes of jet emission are calculated from the dependent trend between the magnitude and the color index at the saturation stage (Equation~\ref{satci}). The I band magnitudes are calculated with the corresponding color indices. Then the magnitudes of V band and I band are converted into fluxes, and add the fluxes of the disk emission of V band and I band, respectively. Finally, we calculate the magnitudes and the color indices with the combined flux. The disk fluxes of V and I bands are extrapolated from the luminosity at 1350 \AA~with a powerlaw index. In order to take into account the uncertain disk features of 3C 454.3, we consider multiple values of disk luminosity and spectral index. The disk luminosities at 1350 \AA~and the spectral indices are varied from 0.5 to 4.0 $\times 10^{46}$ (with step 0.1 less than 1.0 and step 1.0 for larger than 1.0) and -1.0 to 1.0 (with step 0.1), respectively. The simulated magnitudes and color indices for the mixture of the disk and jet emission are plotted in the left panel of Figure~\ref{ci} (different dotted lines correspond to different values of disk luminosity and spectral index).

For disk luminosities fainter than $10^{46}$ erg s$^{-1}$ (black dotted lines in Figure~\ref{ci}), the influence of the disk radiation is hardly observed when the source is brighter than about 17 mag. For disk luminosities brighter than $10^{46}$ erg s$^{-1}$ and spectral indices larger than 0.3 (the magenta dotted lines), the predicted colors are much redder than the observed ones. The results for disk luminosities brighter than $10^{46}$ erg s$^{-1}$ and spectral indices less than 0.3 (the green dotted lines) also have large deviations from the observations. We also consider the condition that the color index of the jet emission is constant~\citep[$\alpha = 1.5$, ci$ = 1.06$]{2013ApJ...773..147J}. The results are plotted in the right panel of Figure~\ref{ci}. With such low disk luminosity, the redder-when-brighter trend is hardly observed at the brightness less than about 16 mag. More importantly, the predicted colors are much redder than the observed ones.

According to the results of the simulations, it seems difficult to produce the observed color variability only considering the contamination of the disk radiation. The combined emission of jet and disk can produce a redder-when-brighter trend, but this trend should be obvious at fainter state for the observed disk luminosity of 3C 454.3 (Figure~\ref{ci}). Thus in despite of the contamination of the disk radiation, the two-stage trend of 3C 454.3 requires that the jet emission contains more than one trend between brightness and color index. Considering the similar trend (redder-when-brighter for faint state and saturation effect for bright state) at long term lightcurve~\citep{2008A&A...491..755R, 2015NewA...36...19Z}, it can be expected that high and low states have different dependent trends between brightness and spectral slope, which may indicate two distinct variability origins, or different components~\citep{2013ApJ...773..147J}. During the outburst, the variability origin is mainly related to the knot ejection~\citep{2013ApJ...773..147J}. For the other relatively small flares without corresponding knot ejections, the flux variability may origin from the variation of other factors (see the discussions in Section 3.4).

\citet{2012ApJ...756...13B} built magnitude-color diagram of 3C 454.3 with J band magnitudes and B-J colors. Their results only showed a redder-when-brighter trend without the saturation stage. For the entire data set of SMARTS monitoring, it shows a slight saturation effect~\footnote{http://www.astro.yale.edu/smarts/glast/tables/3C454.tab}. The transition magnitude is close to 12.0 mag at J band (There are few data points brighter than 12.0 mag in~\citealt{2012ApJ...756...13B}). The disk brightness is 15.40 mag at J band (for disk luminosity $4.0 \times 10^{46}$ erg s$^{-1}$ at 1350 \AA~and $\alpha = 0.5$). The difference is  much larger than that of V band (15.15 mag compared with 16.73 mag). If the contamination of disk emission is the unique reason for the redder-when-brighter trend of 3C 454.3, this trend would get flatter at much fainter state than 12.0 mag for J band. The deviation with observation also supports our conclusion from the simulations.

\citet{2017ApJ...844..107I} analyzed the color variability of 3C 279 over long term timescale. They concluded that a FSRQ can go through three stage of color variability for combined jet and disk radiation in generally, redder-when-brighter, constant color and bluer-when-brighter. For 3C 454.3, the former two stages are obvious for long timescale~\citep{2008A&A...491..755R, 2015NewA...36...19Z}. The bluer-when-brighter trend is slightly observed at bright state in 2007~\citep{2008A&A...491..755R, 2011A&A...531A..90Z}. However, our results show no evidence for bluer-when-brighter trend during the Dec. 2009 outburst of 3C 454.3 (Althrough the color during the brightest flare seems slightly bluer than other small flares. Figure~\ref{lc3} panel g). On the contrary, $\gamma$-ray variability shows slightly harder-when-brighter trend during the outburst (Figure~\ref{lc3} panel f, also see~\citealt{2010ApJ...721.1383A, 2011ApJ...733L..26A}). The different spectral behaviors at the bright days of 3C 454.3 may be related to different electron acceleration mechanisms (also see the discussions in Section 3.4)

\subsection{The flux correlation between optical and $\gamma$-ray}
\begin{figure}
\begin{center}
\includegraphics[angle=0,scale=.35]{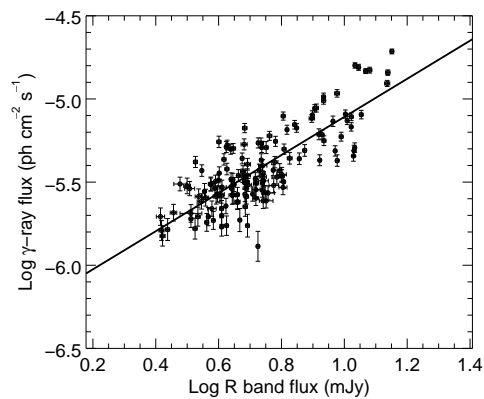}
\caption{The correlation between optical and $\gamma$-ray flux. The solid line shows the best fit. \label{gr}}
\end{center}
\end{figure}

Figure~\ref{gr} presents the flux correlation between optical R band flux (R band magnitude is converted to flux with the zero point 3064 Jy,~\citealt{1998A&A...333..231B}) and the integral $\gamma$-ray flux from 0.1 to 300 GeV during the Dec. 2009 outburst. The solid line gives the best linear fit with
\begin{equation}
 \log F_{\gamma} = (1.14\pm0.07)\log F_R-(6.25\pm0.05) .
\end{equation}

The flux of synchrotron radiation is dependent on the number of the emitting electrons. For $\gamma$-ray emission produced by synchrotron-self Compton (SSC) process, if only the electron number density varies during the flux variability, the $\gamma$-ray emission would vary quadratically with optical emission (mainly dominated by synchrotron radiation at high state) $F_{\gamma} \propto F_R^2$. For EC process, this relation has the form $F_{\gamma} \propto F_R$~\citep{2011MNRAS.410..368B}.~\footnote{There are two main processes to generate the high energy emission from hadronic process, synchrotron radiation of relativistic protons and the p$\gamma$ process~\citep{2013ApJ...768...54B}. Both processes are difficult to fit the break spectra at GeV of some FSRQs and require extreme power of protons~\citep{2013ApJ...768...54B}. Thus we do not consider hadronic model in this work.} Our result is consistent with the prediction of EC process. Moreover, this relation indicates that the variation of electron number density (e.g., electron injection) is important for generating the outburst. This is also supported by the VLBI observations, where new knots emerge during the $\gamma$-ray outbursts~\citep{2001ApJ...556..738J, 2013ApJ...773..147J}.

\citet{2011MNRAS.410..368B} used the \textit{Swift} UVOT data and found $F_{\gamma} \propto F_{UV}^{1.57}$ for 3C 454.3. The deviation is caused by the obvious disk radiation at UV band~\citep{2007A&A...473..819R}, especially at the low state. Only when the jet is bright and the jet emission is dominant (such as the R band in our case), the relation predicted by the EC process could be obvious.

\subsection{The flux ratio between $\gamma$-ray and optical}
The brightest fluxes of optical and $\gamma$-ray occurred on the same day (MJD 55167, hereafter the flare around MJD 55167 is called major flare) during the 2009 outburst. Figure~\ref{lc3} (h) shows the flux ratio between $\gamma$-ray and optical $F_{\gamma}/F_R$ of this outburst. As expectation from the flux correlation $F_{\gamma} \propto F_R$ derived above, $F_{\gamma}/F_R$ is generally a constant. However, there are also several small variations. An increase of $F_{\gamma}/F_R$ is shown at the beginning of the major flare. Then it declines when the flux falls during the major flare. At the end of our observation of this outburst, $F_{\gamma}/F_R$ also shows a declining trend along with the flux falling.

The flux ratio $F_{\gamma}/F_R$ can be taken as the Compton dominance if the slope of SED is invariable over time. Thus $F_{\gamma}/F_R \sim CD \propto \delta^2 u_{ext}/u_B$ for EC process (where $CD$ is Compton dominance, $u_{ext}$ is the energy density of the external photon field, $u_B = B^2/8\pi$ is the energy density of the magnetic field, $\delta$ is the Doppler factor, see e.g.,~\citealt{2013ApJ...763..134F}). That means $F_{\gamma}/F_R$ would be constant if no other parameter changes, except for the electron number density during the flux variability. Obviously, this is not the case in observation. Therefore, it can be expected that there are some other parameters varying during the outburst (changed spectra or others, see the discussions in next section).

The general variability profile can be a result of changing Doppler factors. If this is the case, it would result in the same trend between lightcurve and $F_{\gamma}/F_R$ under the EC process for homogeneous jet (as $F_{\gamma}/F_R \sim CD \propto \delta^2$), which is not seen in our results. This indicates that the variation of the Doppler factor is not the main reason producing the variability, which is also consistent with the scenario indicated by the flux correlation.~\citet{2013ApJ...773..147J} compared the SEDs for three different outbursts and the quiescent state. They found that for the outburst with larger variability amplitude, Doppler factor was also larger than others. Thus the variation of the Doppler factor is still important for different variability amplitude of various outbursts.

\subsection{The detailed structures in optical and $\gamma$-ray lightcurves}
\begin{figure}
\epsscale{0.5}
\plotone{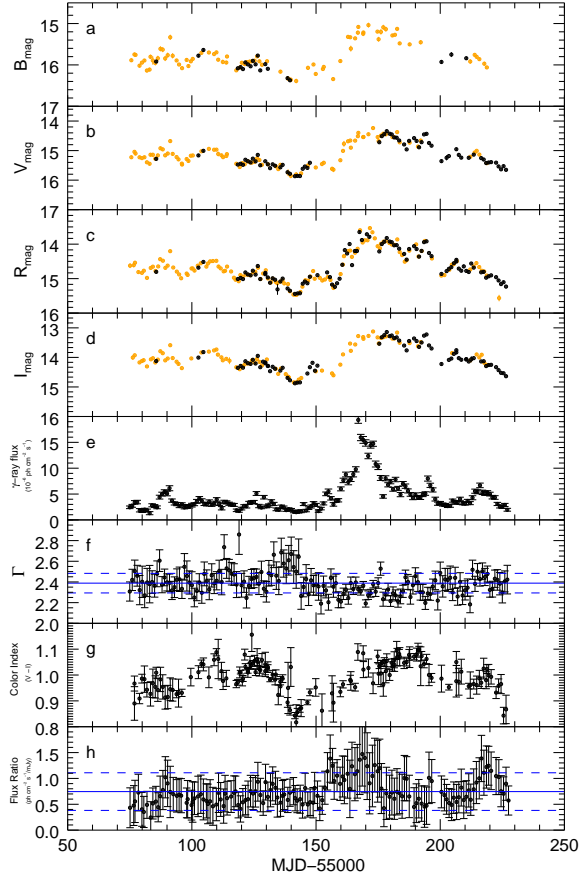}
\caption{The lightcurves of Dec. 2009 outburst. From top to bottom are optical magnitudes of B, V, R and I, Fermi/LAT $\gamma$-ray flux of 0.1-300 GeV, the $\gamma$-ray photon spectral index $\Gamma$ of single powerlaw, the color index V - I and the flux ratio $F_{\gamma}/F_R$, respectively. The orange points in panel (a) to (d) represent the data taken from WEBT. In panel (f) and (h), the mean values of $\Gamma$ and $F_{\gamma}/F_R$ are labelled by blue solid lines. The blue dashed lines in panel (f) and (h) show the mean errors of $\Gamma$ and $F_{\gamma}/F_R$, respectively. \label{lc3}}
\end{figure}

\begin{figure}
\begin{center}
\includegraphics[angle=0,scale=.35]{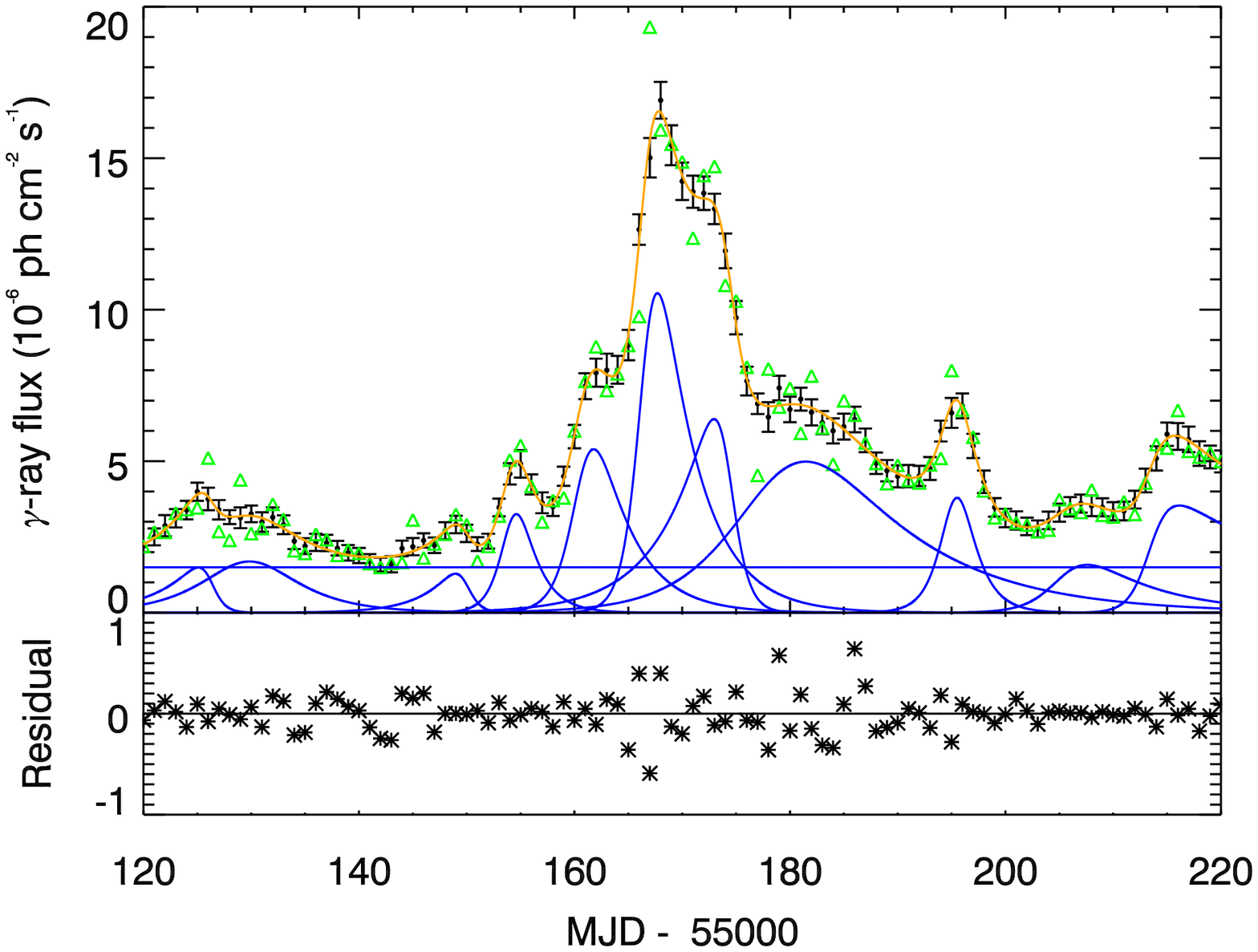}
\includegraphics[angle=0,scale=.35]{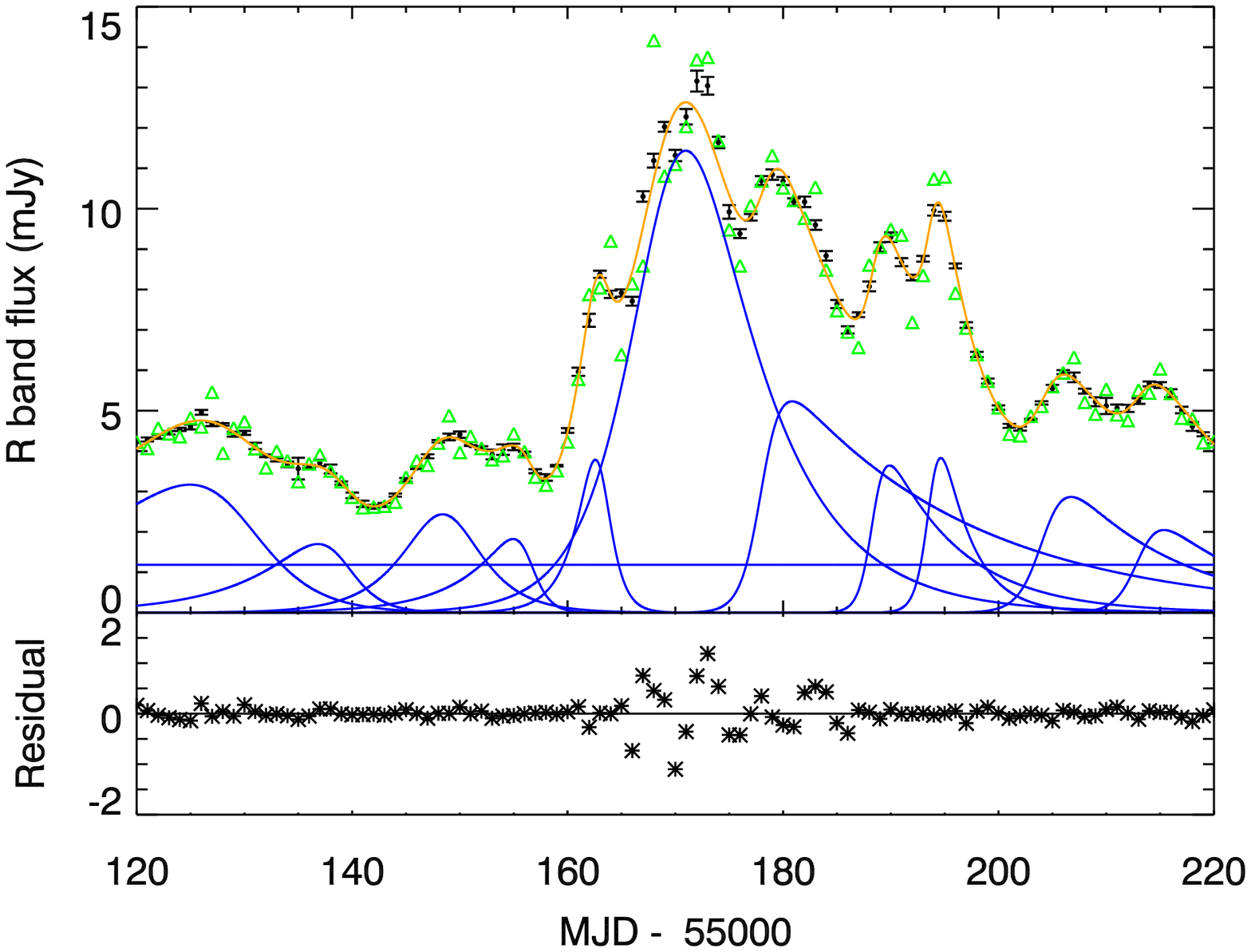}
\caption{The decomposition of lightcurve. left: $\gamma$-ray. Right: optical. The top panel shows the fit, where the green triangles represent the data, the black points are the fitting data after interpolated and smoothed. The blue lines show the baseline and the components of the 11 exponents. The orange line is the total flux of the fitting. The bottom panel shows the residuals. \label{lcfit}}
\end{center}
\end{figure}

There are some differences in the detailed variability behaviors between optical and $\gamma$-ray bands. In particular, the R band flux declines slower than the $\gamma$-ray flux. After the major flare at MJD 55167, it seems that there are several minor flares following at R band, which have no direct counterparts at $\gamma$-ray band (Figure~\ref{lc3}). In order to study these differences in details, we perform a decomposition for the lightcurves with multiple exponents. Each component has the form
\begin{equation}
 F = 2 F_0 (e^{(t_0-t)/T_r}+e^{(t-t_0)/T_f})^{-1},
 \label{exp}
\end{equation}
where $T_r$ and $T_f$ are the timescales for rising and falling, respectively, $F_0$ is the flux at $t_0$~\citep{2011ApJ...733L..26A}. The R band lightcurve is linearly interpolated firstly to match the daily time resolution of the $\gamma$-ray data. Then the lightcurves are smoothed to reduce the random variations. Before fitting, the minimum values of optical and $\gamma$-ray flux are subtracted from the data, respectively. Each lightcurve requires 11 components of Equation~\ref{exp}. The results are plotted in Figure~\ref{lcfit}. The fitting parameters of each component are listed in Table~\ref{fit}.

The sub-flares of $\gamma$-ray generally correspond to those of optical. The biggest difference between the structures of two lightcurves occurs after the major flare. Two minor flares ($t_0$ = MJD 55178.1 and 55188.4, Table~\ref{fit}) follow the major flare at optical without $\gamma$-ray counterparts. On the other hand, the $\gamma$-ray flux drops rapidly with two plateaus ($t_0$ = 55174.2 and 55179.1, Table~\ref{fit}). The first plateau starts approximately 4 days after the brightest flux (the black points in Figure~\ref{lcfit}). The second plateau seems corresponding to the two minor flares at optical. These different structures on the lightcurves are difficult to explain by simple electron injection in the emission region(s).

There are some possible explanations for the orphan flare at optical band. A natural one is that it comes from another component, e.g., thermal radiation from accretion disk. However, the emission lines show rare variability except two flares corresponding to $\gamma$-ray flare, which indicates that disk radiation is slow varying~\citep{2013ApJ...779..100I}. The increase of magnetic field strength would only increase the flux of synchrotron and SSC components, but not increase that of EC component. Meanwhile, the increase of magnetic field strength leads to a decrease of the Compton dominance, which is actually seen from the variation of $F_{\gamma}/F_R$ (Figure~\ref{lc3}, panel h). However, after the minor flares, $F_{\gamma}/F_R$ still remains low, which seems contradictory to the variations of magnetic field strength.

Increasing the energy density of the external field can result in the increase of $F_{\gamma}/F_R$. A flare of emission line has been observed during the 2009 outburst~\citep{2013ApJ...779..100I}. But this flare is more likely to be associated with the excitation of jet rather than disk (\citealt{2013ApJ...763L..36L, 2013ApJ...779..100I}, but see~\citealt{2011ApJ...736..128P} for the indirect evidence of increasing accretion rate). There is also a possibility that the emission region get closer to the external photon field, which makes the energy density of the external field increasing. But changing the external fields is difficult to explain the orphan flare at optical band. We note that the variation of the external field strength can be important on long term timescale, such as for different outbursts. Because the variability amplitude of $\gamma$-ray can be much larger than that of optical~\citep{2011A&A...534A..87R}.

Another possibility is related to the helical movement of the emission region. As the electron cooling for optical and $\gamma$-ray could be non-simultaneous in the emission region, the observed emission of these two bands may generate at different locations, which correspond to different viewing angles in the helical jet. Then the Doppler factors are different for different wavelengths, which could result in different fluxes for optical and $\gamma$-ray. Moreover, the variation of the Doppler factor can also explain the variation of $F_{\gamma}/F_R$.

In addition, if the injected electron energy distribution (break energy or spectral index) changes for different flares, the SEDs would change accordingly. As the $\gamma$-ray flux is integral between 0.1 and 300 GeV, if spectral index of the minor flare is steeper than that of the major flare, the $\gamma$-ray flux of the former may vary less than the later. The optical flux is little affected by this effect. This effect also affects the flux ratio $F_{\gamma}/F_R$, which results in that $F_{\gamma}/F_R$ can not simply represent Compton dominance (see~\citealt{2012RAA....12.1475F} for the similar behavior of the ratio between $\gamma$-ray and radio luminosity). Meanwhile, this would predict a softer spectrum for the minor flare. Actually, the optical color during the major flare appears to be bluer (smaller color index) than the minor flare (Figure~\ref{lc3}, panel g), which may indicate the possible steeper electron spectrum for the minor flare. However, we do not find obvious differences on $\gamma$-ray spectral index between major and minor flares (Figure~\ref{lc3}, panel f).

As the blob moves to downstream, it collides with the previously ejected blobs. This process generates internal shock and accelerates the electrons. The collision can occur more than once, then produce the multiple flares in the observations. In addition, the magnetic reconnection process can also accelerate electrons and produce the flares. The electron spectrum accelerated by magnetic reconnection process can be much harder than by shocks~\citep{2014ApJ...783L..21S, 2014PhRvL.113o5005G}. Different flares coupled with different spectral behaviors during the outburst may be produced by different acceleration mechanisms. Some brightest flares might be related to magnetic reconnection events (\citealt{2011ApJ...726...90Z, 2016Galax...4...12S}), which lead to harder electron spectra (also bluer-when-brighter trends at optical). Other flares with relatively steeper spectra (corresponding to redder-when-brighter trends), on the other hand, may be produced by shocks.

The discussions above are mainly based on the one-zone leptonic model, which is wildely applied to 3C 454.3 (e.g.~\citealt{2010ApJ...716L.170P, 2011MNRAS.410..368B}). In addition,~\citet{2011A&A...534A..87R} explained the variability of 3C 454.3 as a geometrical effect of changing viewing angles under a curved inhomogeneous jet model (also see~\citealt{1999A&A...347...30V, 2017Natur.552..374R}). Under this model, the emission regions are different for different bands, and the curved jet makes the viewing angles different for different bands. This effect results in different Doppler factors between optical and $\gamma$-ray. The general profile of the lightcurves can be explained by the varying viewing angles over time, which do not require the electron injection. Similar to the helical movement of the emission region, this model can explain the difference on the detailed variability behaviors between optical and $\gamma$-ray, as well as the variation of the flux ratio $F_{\gamma}/F_R$.

~\citet{2013ApJ...773..147J} analyzed the multi-wavelength variability, as well as the behaviors of the mm-wave core and optical polarization of 3C 454.3. They explained the three-peak structure during the outburst as different locations of the emission region. In addition, they presented three more complicated models beyond one-zone leptonic model, namely the recollimation shocks and the turbulent extreme multi-zone (TEMZ) model~\citep{2014ApJ...780...87M}, mini-jet model~\citep{2009MNRAS.395L..29G}, and current-driven instability (CDI,~\citealt{2012MNRAS.427.2480N}), to explain the flux and polarization behaviors along with the evolution of the mm-wave core. TEMZ model seems difficult to fit the X-ray spectrum of 3C 454.3. The emission region of mini-jet and CDI models might be different. The CDI is most prominent at the end of the acceleration and collimation zone of jet~\citep{2013ApJ...773..147J}, while the locations of mini-jets are related to the processes triggered the magnetic reconnection events~\citep{2013MNRAS.431..355G}. In addition, both models require strongly magnetized jet, while the magnetization in the dissipation region of blazar is still under debate~\citep{2015MNRAS.449..431J, 2016Galax...4...12S}. More data with evenly sampling and systematic analyses for the common characteristics (such as the spectral behaviors, the location of emission region, the polarization variability and the timescales) during blazar variability may be helpful to distinguish all these models.

\section{Conclusions}
In this paper, we present our optical monitoring data of 3C 454.3 at Yunnan observatories from Sep. 2006 to Nov. 2011. Based on the multi-color photometry, we explore the origin of the optical color variability of 3C 454.3, which is characterized by a two-stage trend. Thanks to the good cadence for the 2009 outburst of 3C 454.3, we further analyze the correlated variability behaviors of optical R band along with the Fermi/LAT $\gamma$-ray data.

The optical color indices show obvious redder-when-brighter trend when the source is fainter than 15.15 mag at V band, and a clear saturation effect when the source is brighter than 15.15 mag. We perform simulations with multiple disk luminosities and spectral indices to evaluate the impact from disk radiation on the magnitude-color relation. The simulations show that the contamination of the disk radiation alone is difficult to explain the observed color variability, which indicates two distinct components or variability origins at high and low states of 3C 454.3.

We find that the variation of the optical and $\gamma$-ray fluxes follows the relation $F_{\gamma} \propto F_R$, which gives an observational evidence for EC process of the $\gamma$-ray emission. Meanwhile, this relation implies that the main mechanism for the variability is electron injection. We also explore the variation of the flux ratio between $\gamma$-ray and optical, as well as the detailed structures of the lightcurves at both bands. The flux ratio $F_{\gamma}/F_R$ shows small variations during the outburst. There are two minor flares following the major flare, which are only observed at optical band. Based on the one-zone leptonic model of EC process, some possible mechanisms, including the variations of the slope of electron spectrum, magnetic field strength, external field and Doppler factor are discussed. The varying Doppler factors in helical (or inhomogeneous) jets, and changed electron spectra for different flares seem feasible to interpret the observations.

\acknowledgments
The authors thank the anonymous referee for
useful comments that greatly improve the
manuscript. We thank Xin-Wu Cao and Bing Zhang for useful suggestions. We are grateful to all the peoples who participated in the observations on this object. We also acknowledge the support of the staff of the Lijiang 2.4 m and Kunming 1 m telescope. Funding for the telescopes has been provided by CAS and the People¡¯s Government of Yunnan Province. This work is also based on data taken and assembled by the WEBT collaboration and stored in the WEBT archive at the Osservatorio Astrofisico di Torino - INAF (http://www.oato.inaf.it/blazars/webt/). This research is supported by the
National Natural Science Foundation of China (NSFC; grants 11233006, U1431123, U1431228, 11573009, 11622324, 11703093 and 11703077), National Key Program for Science and Technology Research and Development
(Grant 2016YFA0400701) and the CAS (grant KJZD-EW-M06 and QYZDJ-SSW-SYS023). HTL is thankful for the Youth Innovation Promotion Association of CAS. KXL thanks the Chinese Western Young Scholars Program and the Light of West China Program provided by CAS (grant No. Y7XB016001). The work of DHY is also supported by the CAS Light of West China Program.

\bibliographystyle{aasjournal}
\bibliography{bib}

\clearpage

\clearpage

\begin{deluxetable}{ccccc}
\tabletypesize{\scriptsize}
\tablecaption{The daily averaged magnitude of 3C 454.3 observed at Yunnan Observatories \label{data}}
\tablewidth{0pt}
\tablehead{
\colhead{MJD} & \colhead{mag} & \colhead{error} & \colhead{filter} & \colhead{instrument}}
\startdata
53999.608 &  16.421  & 0.054  &  B  & 1m \\
53999.573   &  15.903  &   0.026 &  V  &  1m \\
54069.604   &  15.794  &   0.029   & R & 1m \\
53999.582   &  14.939   &  0.014   &  I  &  1m \\
\enddata
\tablecomments{Table~\ref{data} is published in its entirety in the
electronic edition of the {\it Astrophysical Journal}.  A portion is
shown here for guidance regarding its form and content.}
\end{deluxetable}

\begin{deluxetable}{ccccc}
\tabletypesize{\scriptsize}
\tablewidth{0pt}
\tablecaption{The fitting parameters of the lightcurve decomposition  \label{fit}}
\tablehead{
\colhead{$F_0$} &  \colhead{$t_0$ (MJD-55000)}  & \colhead{$T_r$ (day)}      &
\colhead{$T_f$ (day)} & \colhead{$F_{total}$}
}
\startdata
$\gamma$-ray (10$^{-6}$ ph cm$^{-2}$ s$^{-1}$)  & & &  &  (10$^{-6}$ ph cm$^{-2}$ s$^{-1}$) \\
   1.135  &  126.122   &   3.966  &    0.690   &   7.414 \\
   1.688  &  129.730    &  3.867    &  4.083    & 20.034 \\
   1.047  &  149.787   &   3.001  &    0.711    &  6.681 \\
   3.144  &  154.247  &    1.069    &  1.827   &  14.531 \\
   4.798  &  160.831   &   1.251    &  3.508   &  37.815 \\
   8.916  &  166.686   &   1.022   &   3.650    & 70.500 \\
   5.040  &  174.172  &    4.529   &   0.928  &   47.898 \\
   4.652 &   179.110   &   4.483   &   9.697  &  105.620 \\
   3.764  &  195.331   &   1.290   &   1.654  &   17.466 \\
   1.308  &  205.657   &   1.773    &  6.978  &   17.214 \\
   2.225  &  213.529   &   1.018   &  15.623  &   24.020 \\
\hline
optical (mJy) &  &  &  &  (mJy)  \\
   2.581 & 129.000 &  14.727  &  3.485 &  39.813 \\
   1.433 & 138.675 &   6.655  &  1.845 &  19.489 \\
   2.428 & 148.636 &   3.525  &  3.047 &  25.092 \\
   1.433 & 156.249 &   4.862  &  0.978 &  14.595 \\
   3.613 & 162.988 &   1.889  &  1.010 &  16.817 \\
  10.937 & 169.508 &   3.695  &  6.809 & 184.132 \\
   3.229 & 178.117 &   1.009  & 17.593 & 103.665 \\
   2.468 & 188.443 &   0.685  &  6.882 &  34.084 \\
   3.146 & 193.845 &   0.708  &  2.877 &  19.314 \\
   1.997 & 204.372 &   1.219  & 10.588 &  33.374 \\
   1.433 & 213.199 &   1.131  &  9.572 &  14.472 \\
\enddata
\tablecomments{$F_{total}$ is the total flux of each sub-flare.}
\end{deluxetable}

\end{document}